\documentclass[11pt,letterpaper]{article}

\usepackage{geometry}
\usepackage{bbm}
\usepackage[toc,page]{appendix}
\usepackage{amsmath, amsthm, amsfonts, amssymb}
\usepackage{mathrsfs} 
\usepackage{hyperref}
\usepackage{cleveref}
\usepackage{slashed}
\usepackage{dsfont}
\usepackage{wrapfig}
\usepackage{minibox}
\usepackage{graphicx}
\usepackage{graphics}
\usepackage{epsfig}
\usepackage{xcolor}
\usepackage{array}
\usepackage{tikz}
\usepackage{tikz-cd}
\usetikzlibrary{arrows,calc}
\usepackage{relsize}
\usetikzlibrary{decorations.pathmorphing,calc}
\usepackage{authblk}
\usepackage[labelfont=bf]{caption}
\usepackage{caption}
\usepackage{subcaption}
\usepackage{placeins}
\usepackage{braket}
\usepackage{feyn}
\providecommand{\href}[2]{#2}

\def\b{\beta}
\def\g{\gamma}
\def\G{\Gamma}
\def\d{\delta}
\def\D{\Delta}
\def\ve{\varepsilon}
\def\m{\mu}
\def\n{\nu}
\def\l{\lambda}

\def\s{\sigma}
\def\thintablerule{\hrule height0.4pt}

\def\bm{\bold}

\def\pa{\partial}

\newcommand{\be}{\begin{equation}}
\newcommand{\ee}{\end{equation}}
\newcommand{\bea}{\begin{eqnarray}}
\newcommand{\eea}{\end{eqnarray}}
\newcommand{\bal}{\begin{aligned}}
\newcommand{\eal}{\end{aligned}}
\newcommand{\eq}[1]{Eq.~(\ref{#1})}

\numberwithin{equation}{section}

\newcommand{\nbad}{\begin{equation*} \begin{aligned}}
\newcommand{\nead}{\end{aligned} \end{equation*} }
\newcommand{\bad}{\begin{equation} \begin{aligned}}
\newcommand{\ead}{\end{aligned} \end{equation}}
\numberwithin{equation}{section}
\newcommand{\nl}{\nonumber \\}

\def\thintablerule{\hrule height0.4pt}

\def\bm{\bold}

\def\pa{\partial}
\def\tk{\bm k}
\def\tmk{{|\bm k|}}

\def\o{\omega}

\allowdisplaybreaks

\begin{document}

\tikzset{
    photon/.style={decorate, decoration={snake}, draw=black},
    electron/.style={draw=black, postaction={decorate},
        decoration={markings,mark=at position .55 with {\arrow[draw=black]{>}}}},
    gluon/.style={decorate, draw=black,
        decoration={coil,amplitude=4pt, segment length=5pt}} 
}

\centerline{\Huge Ising Cosmology}

\vskip 1 cm
\centerline{\large Nikos Irges$^{\spadesuit}$\footnote{e-mail: irges@mail.ntua.gr}, Antonis Kalogirou$^{\spadesuit}$\footnote{e-mail: akalogirou@mail.ntua.gr}
and Fotis Koutroulis$^{\clubsuit}$\footnote{e-mail: fotis.koutroulis@fuw.edu.pl}}
\vskip 1cm
\centerline{\it $\spadesuit$. Department of Physics}
\centerline{\it School of Applied Mathematical and Physical Sciences}
\centerline{\it National Technical University of Athens}
\centerline{\it Zografou Campus, GR-15780 Athens, Greece}
\vskip .5cm
\centerline{\it $\clubsuit$. Institute of Theoretical Physics, Faculty of Physics,}
\centerline{\it University of Warsaw, Pasteura 5, PL 02-093, Warsaw, Poland}

\vskip 2.2 true cm
\thintablerule
\vskip 2.0ex

\centerline{\bf Abstract}
Using arguments from holography we propose that the deviation of the cosmological spectral index $n_S$ 
of scalar fluctuations from unity may be controlled almost entirely
by the critical exponent $\eta$ of the $d=3$ Ising model.

\vskip 1.0ex\noindent
\vskip 2.0ex
\thintablerule

\newpage

\pagebreak


\section{Introduction}

In this note we demonstrate that the observed deviation from scale invariance 
of the Cosmic Microwave Background as encoded in the spectral index $n_S$ of scalar fluctuations, 
can be explained by dS/CFT arguments. 
The set up is a real scalar $\phi$ in de Sitter (dS) space with an ``out" observer sitting at its horizon and looking at 
the time where modes are exiting. This can be any time except from the time that defines the horizon for thermal effects to be observable.
The observer sees a scale invariant thermal state $\ket{\rm out;\b_{\rm dS}}$ where $\b_{\rm dS}=1/T_{\rm dS}$ and with
$T_{\rm dS}$ the Gibbons-Hawking temperature of dS space. The thermal state is connected to the Bunch-Davies vacuum $\ket{\rm in}$ by a Bogolyubov Transformation.
The breaking of scale invariance is realized when $T<T_{\rm dS}$. 
The effect of the breaking on $n_S$ can be quantified if a certain eigenvalue equation 
is satisfied. 
The associated eigenvalue appears then via standard dS/CFT rules in the 
Callan-Symnazik equation for the boundary operator that couples to the bulk scalar, as the critical exponent $\eta$ of the $d=3$ Ising model.
The result is independent of the precise form of the bulk action and of the details of the breaking mechanism, as long as the breaking is non-zero and small.
Here, we concentrate only on the boundary where the field theory lives.
More details of the (bulk) construction of the thermal scalar in dS space can be found in \cite{FotisAntonis2}.

\section{The spectral index $n_S$ from dS/CFT}

The crucial observation for this note is that the ($\ket{\rm in}$) $\ket{\rm out;\b_{\rm dS}}$ vacuum is expected to correspond to 
a fixed point in the (UV) IR of the dual field theory. Just outside the IR fixed point the bulk state is $\ket{\rm out;\b}$ with $\b>\b_{\rm dS}$.
On the boundary, the IR conformal point can be recognized as the interacting fixed point in the universality class of 
the $d=3$ Euclidean scalar theory with a renormalized Lagrangian 
\be\label{L}
{\cal L} = \frac{1}{2} \partial_i\s\partial_i\s  - {\l}\s^4\, .
\ee
In principle, a mass term should be added too because it
corresponds to a massive field in the bulk but since the final result depends very weakly on it, for simplicity we will keep the field massless.\footnote{
There is a small price for this. Since there is no massive RG trajectory (in dimensional regularization) that connects the Gaussian with the Wilson-Fisher fixed point,
with a non-zero mass term in the Lagrangian it would be easier to argue that the system never reaches exact scale invariance.
A small mass could develop in the classicaly massless theory for example through a generalized version of the Coleman-Weinberg mechanism.}
In the dS/CFT correspondence \cite{Strominger}, a real scalar bulk field $\phi$ (with fluactuation modes $\phi_\tmk$) of dimension $\Delta_-$ in dS space is dual 
to an operator ${\cal O}$ of the boundary CFT of dimension $\Delta_+$. Then bulk and boundary propagators are related by \cite{Maldacena1,Maldacena2}
(the subscripts remind of the momentum conservation $\d$-function which we omit):
\be
\langle\phi_\tmk \phi_{-\tmk}\rangle\sim \frac{1}{\langle{{\cal O}_\tmk}{{\cal O}_{-\tmk}}\rangle} .
\ee
The scalar curvature perturbations $\zeta_\tmk$ and the dS scalar field (of dimension $\D_{\rm cl,-}=0$) perturbation can be in fact connected 
as $\zeta_\tmk = z(\tau) \phi_\tmk$ \cite{Mukhanov, Sasaki,Garriga}, with the $z(\tau)$ factor determined by the (conformal time $\tau$) time-dependent classical background.
As a result, there is a relation between the correlators 
\be
\langle \zeta_\tmk \zeta_{-\tmk}\rangle\sim \langle \phi_\tmk \phi_{-\tmk}\rangle \, ,
\ee
which is gauge-invariant \cite{Maldacena1}.
Recall then the definition of the spectral index using the above identifications:
\bea
n_S-1 &=& 
\frac{\partial}{\partial \ln \tmk} \ln \left(\tmk^3 \langle \zeta_\tmk \zeta_{-\tmk}\rangle\right)=
\frac{\partial}{\partial \ln \tmk} \ln \left(\tmk^3 \langle \phi_\tmk \phi_{-\tmk}\rangle\right)\nonumber\\
&=& 3 - \frac{1}{\langle {\cal O}_\tmk {\cal O}_{-\tmk}\rangle}  \left( \frac{\partial}{\partial \ln \tmk} \langle {\cal O}_\tmk {\cal O}_{-\tmk}\rangle \right)\, 
\eea
and write the Callan-Symanzik equation for a general 2-point function in momentum space as
\bea
\left( \frac{\partial}{\partial\ln\tmk} - \b_\l  \frac{\partial}{\partial\l} + (3 - 2 \D_{\cal O})\right) \langle {\cal O}_\tmk {\cal O}_{-\tmk}\rangle =0\, ,
\eea
where $\D_{\cal O}=\D_+=[\D_{\cal O}]+\G_{\cal O}$ and $\b_\l\equiv\m \frac{\partial\l}{\partial\m}$.
These combine into \cite{Larsen,McNees,Schaar}\footnote{The analysis of \cite{Larsen,McNees,Schaar} recognizes the scalar tilt in the CMB spectrum
as the anomalous dimension of the boundary operator $\s^4$. 
As far as we can tell, there are two versions of their interpretation.
In our notation they can be expressed to leading order in $\b_\l$ as either $n_S=1-2\G_{\s^4}=1-2\frac{\b_\l}{\l}$
or as $n_S=1-2\G_{\s^4}=1-2\frac{\partial\b_\l}{\partial\l}$. 
The first implies a vanishing anomalous dimension and tilt at the fixed point and the second sees it basically related to the critical exponent $\o$.}
\be\label{nsmom}
n_S = 1 - 2\G_{\cal O}- \b_\l  \frac{\partial}{\partial \l} \ln \langle {\cal O}_\tmk {\cal O}_{-\tmk}\rangle\, ,
\ee
where we have used that $[\D_{\cal O}] \equiv \D_{\rm cl,+} = 3 $.
We define the ``total" anomalous dimension $\g_{\cal O}\equiv \m \frac{\partial }{\partial\m} \ln z_{\cal O}$ and
we also have $\g_\s \equiv  \frac{1}{2} \m \frac{\partial }{\partial\m} \ln Z_{\s}$,
the wave function renormalization of $\s$.
In terms of these counterterms, the "operator" anomalous dimension that shifts $[\D_{\cal O}]$ is 
$\G_{\cal O}\equiv - \m \frac{\partial }{\partial\m} \ln \left(Z_\s^{-1}z_{\cal O}\right)\equiv - \m \frac{\partial }{\partial\m} \ln \left(Z_{\cal O}\right)$,
for an operator that contains two $\s$'s.
These definitions imply the relation $\G_{\cal O} = - \g_{\cal O} + 2 \g_\s$. 

Now the boundary operator that couples to a $\zeta_\tmk$ of $\D_{\rm cl,-}=0$ is ${\cal O}= {\Theta}$, the trace of the Ising energy-momentum tensor $\Theta=\d_{ij}{\cal T}_{ij}$,
which being associated with a conserved current, has an exactly vanishing anomalous dimension:
$\G_{\Theta}\equiv 0$.\footnote{The general argument is due to Wilson \cite{Wilson}.
To leading order (or beyond) in the $\ve$-expansion this cancellation can be seen explicitly for example in \cite{Gracey,Claudio,Henriksson}, for ${\cal T}_{ij}$ itself.
For ${\Theta}$, it is realized as a sunset diagram with a $\s\Box\s$ insertion cancelling a usual sunset. 
For all other spin zero operators, 
since $\g_\s\sim O(\ve^2)$ and $\g_{\cal O}\sim O(\ve)$, it is typically taken $\G_{\cal O}\sim - \g_{\cal O}$ to leading order.}
This gives the constraint $\g_{\Theta}= 2\g_\s$. 
In coordinate space we can also write \eq{nsmom} for ${\Theta}$, as
\be\label{nS_CS}
n_S = 1 + \frac{\partial}{\partial \ln\m} \ln \langle {\Theta}(x_1) {\Theta}(x_2)\rangle = 1 - \b_\l  \frac{\partial}{\partial \l} \ln \langle {\Theta}(x_1){\Theta}(x_2)\rangle \, .
\ee
Near the IR fixed point the anomalous dimension $2\g_\s$ is known as the critical exponent $\eta$ and it has a value that
has been computed analytically, among others, in the $\ve$-expansion and numerically on the lattice.
It has the value $\eta\simeq 0.036$ approximately.
Now write the Callan-Symanzik equation in a form where the cancellation of the two contributions to $\G_\Theta=0$ is inserted explicitly and 
with the two cancelling terms shared between the two derivatives:
\be\label{CSmu}
\left[\left(\frac{\partial }{\partial\ln\m} + \eta \right) + \left(\b_\l\frac{\partial }{\partial\l} - \eta \right)\right] \langle {\Theta}(x_1) {\Theta}(x_2)\rangle \simeq 0\, .
\ee
It is tempting to assume that just outside the IR conformal point
the deviation from scale invariance can be parametrized by 
an RG flow, with the two parentheses in the Callan-Symanzik equation
vanishing separately. Then, the scaling equation that the $c_{\Theta}$-coupling approximately satisfies is
$\b_\l {\partial_\l c_{\Theta}}= \eta c_{\Theta}$, where $\langle {\Theta}{\Theta} \rangle = c_{\Theta} /|x|^{2d}$.
The leading order solution is 
\be\label{soleigreg}
c_{\Theta} \sim\left(\frac{16\pi^2-3\l}{\l}\right)^\eta
\ee
and vanishes on the interacting fixed point, as it should. For this reason the eigenvalue equation is non-empty 
only outside the fixed point. Notice that $c_\Theta>0$ between the free and interacting fixed points.

Applying the above eigenvalue to \eq{nS_CS} in the vicinity of this fixed point
and using the non-perturbative value of $\eta$, we are lead to an interesting statement:
\be\label{nS}
n_S \simeq 1 - \eta = 0.964\, .
\ee
The two main sources of errors in this number are the small corrections due to the non-zero $\b_\l$
just outside the fixed point and any errors in the lattice Monte Carlo measurement of $\eta$.

In the remaining we give possible justification for the separate vanishing of the two terms in the parentheses in \eq{CSmu}.
One can give arguments from both the boundary and the bulk perspectives.
On the boundary the argument starts by noticing that the renormalized $\Theta$ is not constructed from the 
bare quantity in the traditional way, since $Z_\Theta=1$. Instead, first $\s_0^4$ is renormalized and $\b_\l$ is computed
and then the renormalized operator is constructed via $\Theta=-\b_\l \m^\ve \s^4$. This leaves the window to construct an RG 
flow by $\Theta=\Theta_0 z^{-1/2}_{\Theta}$. Hitting both sides with $\m \partial / \partial\m$ and using the definitions above,
gives $(\m \partial / \partial\m + \g_\Theta)\langle \Theta \Theta \rangle = 0$. In the bulk, following \cite{Larsen}
one considers the late time equation of motion $\frac{d H}{dt} = - \frac{1}{2} \left(\frac{d \phi}{dt}\right)^2$ in $M_{\rm pl}=1$ units ($H=\frac{d a/dt}{a}$ and $a=e^{Ht}$ is 
the scale factor) and
the exact form of the 2-point function $\langle \Theta_\tmk\Theta_{-\tmk}\rangle \sim \frac{\tmk^3}{H^2}$ and relates the time-dependence
of $H$ away from the conformal limits with an RG flow on the boundary, by the identifications $\m=aH$ and $\l=\phi$.
The latter identification is not sufficient however to reproduce our RG flow because it is not able to see wave function renormalization.
By construction it is sensitive only to $\G_{\cal O}$, therefore it must be generalized.
A possible identification for that purpose is ($\tau$ is the conformal time with $dt=ad\tau$)
\be
\l = \phi - \frac{2\g_\s}{\b_\l} H t \simeq \phi + \ln (H|\tau|)^{\frac{2\g_\s}{\b_\l}}\, ,
\ee
that indeed gives $\left(\b_\l \partial / \partial\l - 2\g_\s + O(\b_\l^2)\right)\langle \Theta \Theta \rangle = 0$ and which near the fixed point reduces to
the eigenvalue equation we need.
These considerations should not be considered of course as a proof but as plausibility arguments.
A rigorous analysis looks substantially more complicated from both sides \cite{Claudio, SkenderisRen}. In both cases the main obstruction seems to be
the disentanglement of the role of wave function renormalization from the renormalization process.

The index in \eq{nS} is just one of many that can be completely fixed by such arguments. In principle also other critical exponents may
be mapped to some cosmological parameters. 
Moreover, the value of $\eta$ fixes more than one observables. A few such examples are given in \cite{FotisAntonis2}.

\section{Tensor fluctuations}

The dual of dS is a non-unitary CFT with negative central charge and the 2-point function of gravitational waves $\g_{\m\n}$
is inversely proportional to the central charge of the dual CFT $P_T\sim \braket{\g\g}\sim \frac{1}{c_T^\star}$ with $c_T^\star\sim -R^2_{\rm dS}$ \cite{Maldacena1} in units $M_{\rm Pl}=1$.
As we have emphasized, the interesting physics happens just outside the scale invariant point where the tensor to scalar ratio
$r = \frac{P_T}{P_S}$ should be small $< O(0.1-0.01)$ and positive.\footnote{The most recent measurement of the tensor to scalar ratio is $r<0.036$ using the latest BICEP/Keck data 
and $r<0.032$ adding also the Planck 2020 data \cite{Planck_r}.}
It is possible to realize a positive $P_T$ (therefore a positive $r$) in the vicinity of a non-unitary conformal point, if the 
effective coupling that determines the physical tensor spectrum is positive 
away from the fixed point and somewhere before it reaches the fixed point turns negative. 
Let us call such a coupling, function of ${\bf e}\equiv x_{12}=|x_1-x_2|$, the $C$-function and denote the IR fixed point by $\star$.
The necessary conditions are then that $C>0$ somewhere outside the fixed point, ${\dot C}\equiv {\bf e} \frac{d C}{d { \bf e}}<0$ and $C^\star=c_T^\star$.

To justify the existence of such an effective coupling we remind \cite{Maldacena1} that the 2-point function of the boundary energy-momentum tensor $\braket{{\bf T}_{\m\n} {\bf T}_{\rho\s}}$ couples to 
the (inverse of) bulk 2-point function $\braket{\gamma_{\m\n} \gamma_{\rho\s}}$ with $\gamma_{\m\n} = \int \frac{d^3k}{(2\pi)^3} \sum_{s=\pm} \epsilon_{\m\n}^s(k) \gamma_{\tk}^s(t) e^{i {\vec x}\cdot {\vec k}}$
and the tensor spectrum is finally governed by the scalar quantity
\be
P_T\sim \braket{\gamma_\tk^s \gamma_{-\tk}^{s'}} = (2\pi)^3 \frac{(\frac{d a}{dt})^2}{\tmk^3} \delta_{ss'}
\ee 
as the tensorial structure carried by the polarization tensor $\epsilon_{\m\n}^s$ gets reduced to a constant factor $\delta_{ss'}$ that sets the helicities $s$ and $s'$ equal.
This implies that there is an effective coupling that one can define, such that
\be\label{Abold}
\braket{{\bf T}_{\m\n} {\bf T}_{\rho\s}} \equiv \frac{{\bf A}_{\m\n\rho\s}({\bf e})}{{\bf e}^{2d}} \sim \frac{C({\bf e})}{{\bf e}^{2d}} F_{\m\n\rho\s}\, ,
\ee
with $F_{\m\n\rho\s}$ a tensor structure independent of ${\bf e}$, whose precise form is irrelevant. Our aim is to define a $C({\bf e})$ with properties that are compatible with observations.

We start from the tensor structure of the most general reducible 2-point function, invariant under the symmetry $e_\m\to -e_\m$ \cite{Karateev}:
\bea\label{TreduCorr}
\braket{{\bf T}_{\m\n} {\bf T}_{\rho\s}} &=& \frac{c_1}{{\bf e}^{2d+4}} e_\m e_\n e_\rho e_\s + \frac{c_2}{{\bf e}^{2d+2}} \left(e_\m e_\n \delta_{\rho\s} + e_\rho e_\s \delta_{\m\n}\right)\nl
&+& \frac{c_3}{{\bf e}^{2d+2}} \left(e_\m e_\rho \delta_{\n\s} + e_\n e_\rho \delta_{\m\s} + e_\m e_\s \delta_{\n\rho} + e_\n e_\s \delta_{\m\rho} \right)\nl
&+&  \frac{c_4}{{\bf e}^{2d}} \delta_{\m\n} \delta_{\rho\s} +  \frac{c_5}{{\bf e}^{2d}} \left(\delta_{\m\rho}\delta_{\n\s} + \delta_{\n\rho}\delta_{\m\s} \right)\, ,
\eea
which defines the tensor ${\bf A}_{\m\n\rho\s}({\bf e})$ in \eq{Abold}.
We have to construct the 2-point functions of the irreducible representations in the decomposition ${\bf T}_{\m\n}=T_{\m\n}+\Theta\frac{\delta_{\m\n}}{d}$ with $T_{\m\n}$ the traceless part.
\begin{itemize}
\item The trace correlator is then
\be\label{ThThgen}
\braket{\Theta\Theta} \equiv \frac{c_\Theta}{{\bf e}^{2d}} = \delta_{\m\n}\delta_{\rho\s} \braket{{\bf T}_{\m\n} {\bf T}_{\rho\s}} \, .
\ee
\item The mixed correlator is
\be\label{TThgen}
\braket{\Theta T_{\rho\s}} \equiv \frac{c_M}{{\bf e}^{2d}} h_{\rho\s} = \braket{\Theta {\bf T}_{\rho\s}} - \braket{\Theta\Theta} \frac{\delta_{\rho\s}}{d}
= \delta_{\m\n} \braket{{\bf T}_{\m\n} {\bf T}_{\rho\s}} - \braket{\Theta\Theta} \frac{\delta_{\rho\s}}{d}\, .
\ee
Indeed, $\delta_{\rho\s} \braket{\Theta T_{\rho\s}} = \braket{\Theta\Theta} - \braket{\Theta\Theta} \frac{\delta_{\rho\s}\delta_{\rho\s}}{d} =  \braket{\Theta\Theta} - \braket{\Theta\Theta} = 0$
as it should, since $T_{\rho\s}$ is traceless. Also, 
\be
\delta_{\rho\s} \braket{\Theta T_{\rho\s}} = 0 \Rightarrow \frac{c_M}{{\bf e}^{2d}} \delta_{\rho\s} h_{\rho\s} = 0 \Rightarrow {\rm Tr}\{h\} = 0\, ,
\ee
with
\be
h_{\m\n} = \frac{\delta_{\m\n}}{d} - \frac{e_\m e_\n}{{\bf e}^2} \, .
\ee
\item The 2-point function of the traceless representation is
\be\label{TTgen}
\braket{T_{\m\n} T_{\rho\s}} \equiv \frac{A_{\m\n\rho\s}}{{\bf e}^{2d}} = \braket{{\bf T}_{\m\n} {\bf T}_{\rho\s}} - \frac{c_M}{{\bf e}^{2d}} h_{\m\n} \frac{\delta_{\rho\s}}{d}
- \frac{c_M}{{\bf e}^{2d}} h_{\rho\s} \frac{\delta_{\m\n}}{d} - \frac{c_\Theta}{{\bf e}^{2d}} \frac{\delta_{\m\n}}{d} \frac{\delta_{\rho\s}}{d}\, .
\ee
It is easy to check that $\delta_{\m\n} \braket{T_{\m\n} T_{\rho\s}} = \delta_{\rho\s} \braket{T_{\m\n} T_{\rho\s}} = 0$.
The $A_{\m\n\rho\s}$ here should not be confused with the ${\bf A}_{\m\n\rho\s}$ of \eq{Abold} which is the tensor structure of the reducible 2-point function.
\end{itemize}
Substituting \eq{TreduCorr} into \eq{ThThgen}, \eq{TThgen} and \eq{TTgen}, we get
\bea
c_\Theta &=& c_1 + 2d c_2 + 4c_3 + d^2 c_4 + 2d c_5\nl
c_M &=& - (c_1 + d c_2 + 4c_3)
\eea
and
\bea\label{Aexpanded}
A_{\m\n\rho\s} &=& \frac{1}{d^2} (c_1 + 4c_3 - 2dc_5) \delta_{\m\n}\delta_{\rho\s} + c_5 (\delta_{\m\rho}\delta_{\n\s} + \delta_{\n\rho}\delta_{\m\s})\nl
&+& \frac{c_3}{{\bf e}^2}  \left(e_\m e_\rho \delta_{\n\s} + e_\n e_\rho \delta_{\m\s} + e_\m e_\s \delta_{\n\rho} + e_\n e_\s \delta_{\m\rho} \right)
+ \frac{c_1}{{\bf e}^{4}} e_\m e_\n e_\rho e_\s \nl
&-& \frac{1}{d {\bf e}^2} (c_1 + 4c_3)  \left(e_\m e_\n \delta_{\rho\s} + e_\rho e_\s \delta_{\m\n}\right)\, .
\eea
Writing the tensorial structure at the fixed point \cite{Petkou} as
\bea
I_{\m\n\rho\s} &=& -\frac{2}{d}  \delta_{\m\n}\delta_{\rho\s} + (\delta_{\m\rho}\delta_{\n\s} + \delta_{\n\rho}\delta_{\m\s})\nl
&-& \frac{2}{{\bf e}^2}  \left(e_\m e_\rho \delta_{\n\s} + e_\n e_\rho \delta_{\m\s} + e_\m e_\s \delta_{\n\rho} + e_\n e_\s \delta_{\m\rho} \right)
+ \frac{8}{{\bf e}^{4}} e_\m e_\n e_\rho e_\s
\eea
and comparing with \eq{Aexpanded}, yields the boundary conditions via $c_T^\star I_{\m\n\rho\s}=A_{\m\n\rho\s}^\star$:
\be\label{starvalues}
c_1^\star = 8 c_T^\star\, , \hskip .3cm c_2^\star =0 \, , \hskip .3cm c_3^\star = -2 c_T^\star \, , \hskip .3cm c_4^\star = - \frac{2}{d} c_T^\star  \, , \hskip .3cm c_5^\star = c_T^\star
\ee
The conservation equations 
\be
\pa_\mu \braket{{\bf T}^{\m\n} {\bf T}^{\rho\s} } = 0\, 
\ee
should be solved with these boundary conditions.
The result in $d=3$ is that the conservation equations are satisfied if
\bea\label{diffsystC}
&& {\dot c}_1 + {\dot c}_2 + 2 {\dot c}_3 - 4c_1 - 8c_2 -16 c_3 = 0 \nl
&& {\dot c}_3 + {\dot c}_5 + c_2 - 3c_3 - 6 c_5 = 0 \nl
&& {\dot c}_2 + {\dot c}_4 - 4c_2 + 2c_3 - 6c_4 = 0\, .
\eea
Unlike in $d=2$ these equations show no obvious preference for a particular $C$-function. 
In our case however we have the constraint of the eigenvalue equation that $c_\Theta$ satisfies, 
written in terms of the ${\bf e} \frac{d}{d {\bf e}}$ derivative as ${\dot c}_\Theta = - \eta c_\Theta$, which forces
the couplings of the five independent tensor structures to satisfy individually the same eigenvalue equation ${\dot c}_i = - \eta c_i$, $i=1,\cdots,5$.
This causes the system of differential equations \eq{diffsystC} to collapse into an algebraic system that can be solved for three of the couplings in terms of the other two.
The solution determines in particular that $\frac{c_M}{c_\Theta}=\frac{6+\eta}{2(3+\eta)}$.
Note that the general solution to the eigenvalue equation in position space is
\be\label{soleig}
c_i({\bf e}) = c_i^\star \left[\frac{\bf e}{{\bf e}_\star}\right]^{-\eta}\, .
\ee
Now we can define the $C$-function as the most general form that can be constructed from the five couplings:\footnote{The solution \eq{soleig} 
is valid when $c_i^\star\ne 0$, otherwise the eigenvalue equation is singular. 
This means that when $c^\star=0$ (in our case for $c_\Theta$, $c_M$ and $c_2$)
the eigenvalue equation should be regularized by either introducing a small cut-off $c_\epsilon^\star << 1$ or by using the Callan-Symnazic equation as in \eq{soleigreg}.
Note also that since both $c_M$ and $c_\Theta$ contain $c_2$ in their decompositions in terms of the $c_i$, the singular behaviour is balanced.}
\be
C = \sum_i (a_i c_i + b_i c_i^\star )\, .
\ee
Substituting \eq{starvalues} and \eq{soleig} this can be written as
\be
C = \left(a  \left[\frac{\bf e}{{\bf e}_\star}\right]^{-\eta} + b \right)c_T^\star\, ,
\ee
with $a=8a_1-2a_3-\frac{2}{3}a_4+a_5$ and $b=8b_1-2b_3-\frac{2}{3}b_4+b_5$.
Two of the conditions that we need are
\bea
C^\star = c_T^\star && \longrightarrow\,\,\, a + b = 1\nonumber\\
{\dot C} < 0 && \longrightarrow\,\,\,  a < 0
\eea
and the third, $C>0$, is satisfied as long as for a given ${\bf e}$
\be
a < a_c \equiv \frac{1}{1 - \left[\frac{\bf e}{{\bf e}_\star}\right]^{-\eta}}\, .
\ee
The flow towards the IR is: ${\bf e}_{\rm UV} \to {\bf e}_{\rm IR} \equiv {\bf e}_\star$ and $a_c\to -\infty$. We see that for any value of $a$ for which the inequality is initially satisfied,
at some point $C$ turns negative. This means of course that $b$ must be non-zero and positive.
We summarize the situation in an RG flow picture:
%
\begin{figure}[!htbp]
\centering
\includegraphics[width=6cm]{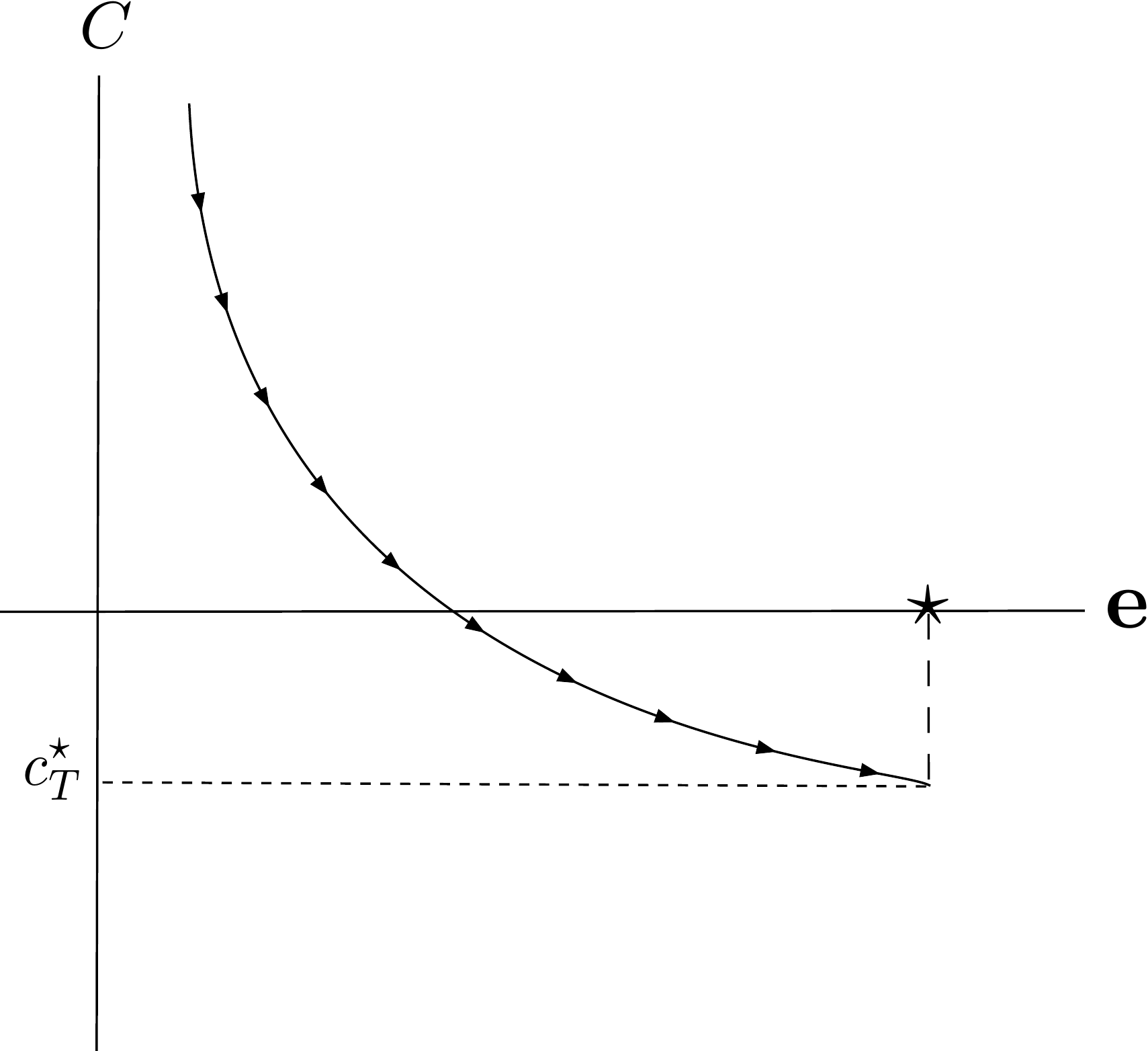}
\end{figure}\label{Cfunction}
\FloatBarrier
%
To compute the actual form of $C$, therefore of $P_T$, is beyond the scope of this letter. 
This would require a demanding calculation from both the field theory and the dS sides. 
From the field theory side the computation of the "renormalized" $\braket{{\bf T}_{\m\n} {\bf T}_{\rho\s}}$ correlator starts from the 2-loop order
in the $\ve$-expansion and it involves the finite parts of the contributing diagrams since $\Gamma_{\bf T}=0$. Such a calculation has not been done to our knowledge.
From the bulk point of view, a simultaneous computation of $P_S$ and $P_T$ needs a knowledge of some sort of time-dependent classical background.
Such a classical background could be either an effective background originating from the late time behaviour of our scalar field or a new
classical field on the top of $\phi$, with the profile of an inflaton.
What we have only demonstrated here is that, in principle, it is possible to have a non-unitary CFT as the dual field theory and still observe a positive 
$r$-index.

Regarding the term "Ising Cosmology" that we introduced here, one could argue that in addition to the fact that the dual of dS is a non-unitary CFT, it should be some large $N$ theory.
The Ising model is unitary and of $N=1$ (in the $O(N)$ critical model sense) thus the anomalous dimension $\eta$ used here one could claim, can not be relevant.
As far as the large $N$ issue is concerned, it is a fact that the anomalous dimension of the basic field is actually the same for any $N$ that has been measured on the lattice
or computed in the large $N$ expansion or by conformal bootstrap. That is, $\eta\simeq 0.036$ is pretty much an $N$-independent number \cite{Henriksson}. 
Now the large $N$ model is unitary and let us assume that it is dual to a system in AdS space.
But the analytic continuation of this AdS theory is the dS-scalar system that we are dealing with here \cite{Maldacena1}.
This means -in one sentence- that the physical state we discuss here which becomes dual to a non-unitary CFT in the scale invariant limit is related via analytic continuation to
an AdS theory whose dual in the scale invariant limit is a large $N$ unitary CFT with a critical exponent $\eta$ that is $N$-independent.
This argument we believe justifies the use of the Ising critical exponent $\eta$, even though strictly speaking the characterization Ising may be an oversimplification.

\vskip .1cm
{\bf Acknowledgments.}
The research of F.K. leading to these results has received funding from the Norwegian Financial Mechanism for years 2014-2021, grant nr DEC-2019/34/H/ST2/00707.
The authors thank A. Kehagias for many valuable discussions and comments and A. Stergiou for a comment.
 


\end{document}